\documentstyle[12pt,epsfig]{article}
\newcommand{\bbox}[1]{\mbox{\boldmath{$#1$}}}
\def\G3H{$\Gamma^{\beta}_t$}
\def\signud{{$\sigma_{\nu d}$}}
\def\signudCC{{$\sigma_{\nu d}^{CC}$}}
\def\signudNC{{$\sigma_{\nu d}^{NC}$}}
\def\AEXC{$\bbox{A}_{\rm EXC}$}
\def\bA{{\bbox{A}}}
\def\signudN{$\sigma_{\nu d}${\scriptsize{(Netal)}}}
\def\signudCCN{$\sigma_{\nu d}^{CC}${\scriptsize{(Netal)}}}
\def\signudNCN{$\sigma_{\nu d}^{NC}${\scriptsize{(Netal)}}}
\def\signudNSGK{$\sigma_{\nu d}${\scriptsize{(NSGK)}}}  
\def\signudCCNSGK{$\sigma_{\nu d}^{CC}${\scriptsize{(NSGK)}}} 
\def\signudNCNSGK{$\sigma_{\nu d}^{NC}${\scriptsize{(NSGK)}}}

\def\bea{\begin{eqnarray}}
\def\eea{\end{eqnarray}}
\title{Neutrino-deuteron reactions at solar neutrino
energies}
\author{
S. Nakamura${}^1$, T. Sato${}^{1,2}$, 
  S. Ando${}^2$, T.-S. Park${}^2$, F. Myhrer${}^2$,
  V. Gudkov${}^2$\\
  and\\
  K. Kubodera${}^2$\\ 
  ${}^1${\it Department of Physics, Osaka University, 
  Toyonaka, Osaka 560-0043, Japan}\\ 
${}^2${\it Department of Physics and Astronomy, 
University of South Carolina,}\\
{\it Columbia, SC 29208, USA}}
\begin{document}
\maketitle
%\draft

\noindent
{\bf Abstract}

In interpreting the SNO experiments,
accurate estimates 
of the $\nu d$ reaction cross sections
are of great importance.
We improve the previous estimates of our group
by updating some of its inputs
and by taking into account the results
of a recent effective-field-theoretical calculation.
The new cross sections are 
slightly ($\sim$1 \%) larger than 
the previously reported values.
It is shown to be reasonable 
to assign 1\% uncertainty to the $\nu d$
cross sections reported here;
this error estimate does {\it not}
include radiative corrections, 
for which we refer to the literature.

\vspace{0.8cm}
\noindent
{\it PACS} : 25.30.Pt, 25.10.+s, 26.65.+t, 95.30.Cq\\
{\it Key words} : neutrino-deuteron reaction, 
solar neutrino, neutrino oscillations,
exchange current, effective field theory
\vspace{0.8cm}

\noindent
{\bf 1.  Introduction}

\vspace{3mm}
The establishment of 
the Sudbury Neutrino Observatory (SNO)
\cite{SNO,ahmad}
has motivated intensive theoretical effort
to make reliable estimates of 
the neutrino-deuteron reaction cross sections
[3-6].
%\cite{kn,KN,NSGK,BCK}. 
One of the primary experiments at SNO
is the measurement of the solar neutrino flux.
By observing the charged-current (CC) reaction, 
$\nu_e d \rightarrow e^- p p$, 
one can determine the flux
of the solar electron-neutrinos 
while, by monitoring the neutral-current (NC) reaction, 
$\nu_x d\rightarrow \nu_x p n$
($x\!=\!e$, $\mu$ or $\tau$),
one can determine 
the total flux of the solar neutrinos of any flavors.
These features make SNO a unique facility
for studying neutrino oscillations.
SNO is also capable of monitoring
the yield of 
the neutrino-electron elastic scattering (ES),
$\nu_e e \rightarrow \nu_e e$,
which also carries information 
on neutrino oscillations.
The first report from SNO \cite{ahmad}
was concerned with the measurements
of the CC and ES processes.
By combining the SNO data on the CC reaction
with the Super-Kamiokande data on
ES \cite{SuperK},\footnote{The SNO data on ES 
is consistent with the Super-K data \cite{SuperK} 
but the latter has higher statistics.}
strong evidence for $\nu_e$ oscillations 
has been obtained \cite{ahmad}.
It is to be noted 
that the sharpness of this important conclusion 
depends on the precision  
of theoretical estimates 
for the $\nu d$-reaction cross sections. 
In the present communication
we wish to describe our attempt to improve
the existing estimates.

We first give a brief survey
of the theoretical estimates of
the $\nu d$-reaction cross section
that were used in the analysis in 
\cite{ahmad}.\footnote{
In what follows, \signudCC and \signudNC
stand for the total cross sections
(in the laboratory frame)
for the CC and NC reactions, respectively;
in referring to \signudCC and \signudNC
collectively, we use the generic symbol,
\signud. 
The incident neutrino energy 
in the lab-frame will be denoted by $E_\nu$.}
A highly successful method for describing 
nuclear responses to electroweak probes
is to consider one-body impulse approximation 
(IA) terms and two-body exchange-current 
(EXC) terms acting on non-relativistic 
nuclear wave functions,
with the EXC contributions derived 
from one-boson exchange diagrams \cite{cr}.
We refer to this method as the standard nuclear 
physics approach (SNPA)\cite{cs}.\footnote{
This approach was called
the phenomenological Lagrangian approach (PhLA)
in \cite{NSGK}.} 
The most elaborate calculation of 
the $\nu d$ cross sections based on SNPA
has been done by Nakamura {\it et al.}\ (NSGK)
\cite{NSGK}.
Since the $\nu d$ reactions 
in the solar neutrino energy ($E_\nu\le$ 20 MeV)
is dominated by the contribution of
the space component, $\bA$,
of the axial current ($A_\mu$),
the theoretical precision of \signud\  is controlled
essentially by the accuracy with which one can
calculate the nuclear matrix element of \bA.
Let us decompose $\bA$ as 
$\bA=\bA_{\rm IA}+\bA_{\rm EXC}$,
where $\bA_{\rm IA}$ and 
$\bA_{\rm EXC}$ are the IA and EXC contributions,
respectively.
Since $\bA_{\rm IA}$ is well known, 
the theoretical uncertainty
is confined to $\bA_{\rm EXC}$.
Now, among the various terms
contributing to \AEXC,
the $\Delta$-excitation current ($\bbox{A}_\Delta$)
gives the most important contribution \cite{CRSW},
and $\bbox{A}_\Delta$ involves the coupling constants
for the $A_\mu N\Delta$ vertex, the $\pi N\Delta$ vertex
and the $\rho N\Delta$ vertex, 
and the corresponding form factors. 
Although the quark model is believed to 
provide reasonable estimates for these coupling constants, 
it is at present impossible to
test their individual values empirically;
only the overall strength of the $\Delta$-excitation 
current can be monitored with electroweak processes 
in a few-nucleon system.
NSGK therefore considered two methods
for controlling the strength of the $\Delta$-excitation current.
In one method, by exploiting the fact 
that the $\Delta$-excitation current features in
the $np\!\rightarrow\!\gamma d$ amplitude as well,
its strength is determined so as to reproduce 
the $np\!\rightarrow\!\gamma d$ cross section.
The second method uses
the tritium $\beta$ decay rate, \G3H, 
and the strength of the $\Delta$-excitation current
is adjusted, as in Refs.\cite{CRSW,model_I}, 
to reproduce the well-known experimental value of \G3H.
The first method was found 
to give \signud\  about 3\% larger than 
the second method,
and NSGK adopted this 3\% difference
as a measure of uncertainty in their calculation
based on SNPA.

Apart from SNPA,
a new approach based on effective field theory
(EFT) has been scoring great success in describing
low-energy phenomena in few-nucleon systems
[12-14].
%\cite{weinberg,epel,EFT}.
Butler {\it et al.}\ (BCK) \cite{BCK} applied EFT
to the $\nu d$ reactions,
using the regularization scheme called
the power divergence subtraction (PDS)
\cite{PDS}.
Their results agree with those of 
NSGK in the following sense.
The EFT Lagrangian used by BCK involves 
one unknown low-energy constant (LEC),
denoted by $L_{1A}$,
which represents the strength of
$A_\mu$-four-nucleon contact coupling.
BCK adjusted $L_{1A}$ to optimize
fit to the \signud\  of NSGK
and found that,
after this optimization,
the results of the EFT and SNPA calculations
agree with each other within 1\%
over the entire solar-$\nu$ energy region.
Furthermore, the best-fit value of $L_{1A}$
turned out to be consistent with what one
would expect from the ``{\it naturalness}" argument
\cite{BCK}.
The fact that the results of 
an {\it ab initio} calculation
(modulo one free parameter) based on EFT are
completely consistent with those of SNPA 
may be taken as evidence 
for the basic soundness of SNPA.

Having given a brief survey of the existing theoretical
estimates of \signud,
we now describe several points
that need to be addressed
for improving the estimates.
We first note that,
as pointed out by Beacom and Parke \cite{beacom},
the value of the axial coupling constant, $g_A$,
used in NSGK is not the most updated one.
This obvious deficiency needs to be remedied.
Secondly, in their treatment of \AEXC,
NSGK left out some sub-dominant diagrams,
and therefore it is worthwhile 
to examine the consequences of using 
the full set of relevant Feynman diagrams
\cite{model_I}.
Furthermore, NSGK adopted as their 
{\it standard run} the case in which 
the strength of the $\Delta$-excitation current was adjusted
to reproduce the measured
$np\!\rightarrow\!\gamma d$ rate.
However, the $np\rightarrow d\gamma$ reaction 
governed by the vector current 
cannot be considered as a better constraint
than \G3H for monitoring
the effective strength of the $A_\mu N\Delta$ vertex  
relevant to the axial-vector transition.  
In the present work, therefore, 
we adopt as our standard choice the case
in which \AEXC\  is controlled by \G3H.
Thirdly, at the level of precision in question,
radiative corrections become relevant
[16-18].
%\cite{beacom,RC_towner,RC_KRV}.
In this communication, however,
we do not address radiative corrections
{\it per se} 
and simply refer to the literature on this issue
\cite{beacom,RC_KRV}.
A related problem is what value should be used 
for the weak coupling constant.
One possibility is to use the standard 
Fermi constant, $G_F$,
which has been derived from $\mu$-decay
and hence does not contain any hadron-related
radiative corrections.
Another possibility is to employ 
an effective coupling constant
(denoted by $G_F^{\,\,\prime}$) that includes the
so-called inner radiative corrections 
for nuclear $\beta$-decay.
NSGK adopted the first choice.
However, since the inner corrections
are established reasonably well,
it seems more natural to use $G_F^{\,\,\prime}$ 
instead of $G_F$.
We therefore adopt here $G_F^{\,\,\prime}$
as the weak coupling constant
(see below for more detail).
An additional point that warrants a further study
is the stability of the calculated value
of \signud\  against different choices
of the $NN$ interactions. 
NSGK investigated this aspect 
for a rather wide variety of modern high-quality
$NN$ interactions \cite{AV18,NN_NSGK}
and found the stability of \signud\ 
at the 0.5\% level. 
The interactions considered in NSGK, 
however, are all local potentials and 
have similar values of the deuteron 
$D$-state probability, $P_D$.
Since the CD-Bonn potential \cite{CD-Bonn}
has a somewhat smaller value of $P_D$
than the other modern high-quality $NN$ potentials,
we study here whether the stability persists
with the use of the CD-Bonn potential.

Besides these improvements within the framework 
of SNPA,
we present here a new comparison between
SNPA and EFT.
Park {\it et  al.}
[22-24]
%\cite{PMR,PKMR,pphep} 
have developed an EFT approach wherein
the electroweak transition operators
are derived with a cut-off scheme EFT
(\`{a} la Weinberg \cite{weinberg}) 
and the initial and final wave
functions are obtained with the use of
the high-quality phenomenological nuclear interactions.
For convenience, we refer to this approach as EFT*.
EFT* applied to the Gamow-Teller transitions
contains one unknown LEC denoted 
by {$\hat{d}_R$}, which plays a role similar to
$L_{1A}$ in BCK.
In EFT*, however, one can determine
{$\hat{d}_R$} {\it directly} from \G3H 
\cite{pphep}.
This allows a parameter-free calculation
of \signud,  
and very recently Ando {\it et al.}  
have carried out this type of 
calculation \cite{ando}.
We present a comparison between our new results 
based on SNPA and those based on EFT*,
and we argue that good agreement between them
renders further support for the robustness
of \signud\ obtained in SNPA.
It will be seen that the new values of \signud\  
are close to those given in NSGK,
but that a significant improvement
in error estimates has been achieved. 

\vspace{4mm}
\noindent
{\bf 2.  Formalism}

\vspace{3mm}
\noindent

\vspace{3mm}
We study the total and differential cross sections for
the CC and NC reactions of neutrinos and antineutrinos with the
deuteron:
\bea
\nu_e +d &&\rightarrow e^-+ p+p \;\;\;\;\;\;\;
({\rm CC})\label{eq_nu_CC}\\
\nu_x +d &&\rightarrow \nu_x +n+p \;\;\;\;\;\;\;
({\rm NC})\label{eq_nu_NC}\\
{\bar{\nu}}_e +d &&\rightarrow e^+ +n+n 
\;\;\;\;\;\;({\bar{\nu}}{\rm -CC})\label{eq_bnu_CC}\\
{\bar{\nu}}_x+d &&\rightarrow {\bar{\nu}}_x+n+p
\;\;\;\;\;\;\;({\bar{\nu}}{\rm -NC})\,,\label{eq_bnu_NC}
\eea
where $x=e, \mu\ {\rm or}\ \tau$.
We briefly describe our calculational framework
and explain in what specific aspects 
we improve upon NSGK\cite{NSGK}.

The four-momenta of the participating particles 
are labeled as
\begin{equation}
\nu/\bar{\nu}(k)+d(P) \rightarrow \ell(k')+N_1(p_1')+N_2(p_2'),
\label{momdef}
\end{equation}
where $\ell$ corresponds to $e^{\pm}$ for the CC reactions
\ [Eqs.(\ref{eq_nu_CC}),(\ref{eq_bnu_CC})], 
and to $\nu$ or $\bar{\nu}$ for the NC reactions
\ [Eqs.(\ref{eq_nu_NC}),(\ref{eq_bnu_NC})].
The energy-momentum conservation reads:
$k+P=k'+P'$ with $P'\equiv p_1'+p_2'$.
A momentum transferred from the
lepton to the two-nucleon system is denoted by 
$q^{\mu} = k^{\mu} - k'^{\mu} = P'^{\mu} - P^{\mu}$.
In the laboratory system, 
which we use throughout this work, 
we write
\begin{equation}
k^\mu=(E_{\nu},\bbox{k}),\,\,
k'^\mu=(E'_{\ell},\bbox{k}'),\,\,
P^\mu=(M_d,\bbox{0}),\,\,
P'^\mu=(P'^0,\bbox{P}'),\,\,
q^\mu=(\omega,\bbox{q}).
\end{equation}

The interaction Hamiltonian 
for semileptonic weak processes is
given by the product of 
the hadron current ($J_{\lambda}$) and 
the lepton current ($L^{\lambda}$) as
\begin{eqnarray}
H_W^{CC} & = & \frac{G_F^{\,\,\prime} V_{ud}}
{\sqrt{2}}\int d\bbox{x}  [
   J_{\lambda}^{CC}(\bbox{x})L^{CC,\lambda}(\bbox{x}) +
    \mbox{h. c.}]\label{eq_Ham-CC}
\end{eqnarray}
for the CC process and
\begin{eqnarray}
H_W^{NC} & = & \frac{G_F^{\,\,\prime}}{\sqrt{2}}\int d\bbox{x}
\  [J_{\lambda}^{NC}(\bbox{x})L^{NC,\lambda}
    (\bbox{x})+\mbox{h. c.}]\label{eq_Ham-NC}
\end{eqnarray}
for the NC process.
Here $G_F^{\,\,\prime}$ is 
the weak coupling constant,
and $V_{ud}$ is the K-M matrix element.
For the weak coupling constant, 
instead of 
$G_F = 1.16637\times 10^{-5}$ GeV$^{-2}$ 
employed in NSGK,
we adopt here
$G_F^{\,\,\prime}=1.1803\times10^{-5}$ GeV$^{-2}$
obtained from $0^+\rightarrow 0^+$ 
nuclear $\beta$-decays \cite{TH}.\footnote{
The relation between $G_F^{\,\,\prime}$
and the quantities used in \cite{TH} is:
$G_F^{\,\,\prime\ 2}=(G_V/V_{ud})^2(1+\Delta_R^V)$,
where $\Delta_R^V$ is the nucleus-independent
radiative correction.}
$G_F^{\,\,\prime}$ subsumes the bulk of 
the {\it inner} radiative corrections.\footnote{
To be precise, the inner corrections 
for the CC and NC reactions may differ
but the difference reported in the literature
\cite{RC_KRV} is comparable to
the estimated uncertainty 
of our present calculation (see below).}
The K-M matrix element is taken to be $V_{ud}$ = 0.9740\cite{TH}
instead of $V_{ud}$ = 0.9749 used in NSGK.

The leptonic currents, 
$L^{CC,\lambda}$ and $L^{NC,\lambda}$,
are well known.
The hadronic charged current is written as
\begin{eqnarray}
J_{\lambda}^{CC}(\bbox{x}) & = & 
V_{\lambda}^{\pm}(\bbox{x}) +
A_{\lambda}^{\pm}(\bbox{x}) ,
\end{eqnarray}
where $V_{\lambda}$ and $A_{\lambda}$ 
denote the vector and axial-vector currents, respectively.
The superscript $+(-)$  denotes 
the isospin-raising (-lowering) operator  for
the $\bar{\nu}(\nu)$-reaction.
The hadronic neutral current is given 
by the standard model as
\begin{eqnarray}
J_{\lambda}^{NC}(\bbox{x}) 
&=& (1-2 \sin^2 \theta_W )V_{\lambda}^{3} +
A_{\lambda}^{3} -2 \sin^2 \theta_W 
V_{\lambda}^{s} , \label{eq_NC-current}
\end{eqnarray}
where $\theta_W$ is the Weinberg angle.
$V_{\lambda}^{s}$ is the isoscalar part 
of the vector current,
and the superscript `3' denotes 
the third component of the isovector current. 
The hadron current consists of 
one-nucleon impulse approximation (IA) terms
and two-body exchange current (EXC) terms.

The IA currents are given in terms of 
the single-nucleon matrix elements of $J_\lambda$.
The standard parameterization for them is
\begin{eqnarray}
<\!N(p')\ |\ V_{\lambda}^{\pm}(0)\ |\ N(p)\!> & = &
  \bar{u}(p')[f_V \gamma_{\lambda} + i \frac{f_M}{2m}
    \sigma_{\lambda\rho}q^\rho]\tau^{\pm} u(p), \\
<\!N(p')\ |\ A_{\lambda}^{\pm}(0)\ |\ N(p)\!> & = &
  \bar{u}(p')[f_A \gamma_{\lambda}\gamma^5 +
  f_P \gamma^5 q_{\lambda} ] \tau^{\pm} u(p)\,,
\end{eqnarray}
where $m$ is the average of the proton and neutron masses.
For the third component of the isovector current, 
we simply replace $\tau^{\pm}$ with $\tau^3/2$. 
The isoscalar current is given as
\begin{eqnarray}
<\!N(p')\ |\ V_{\lambda}^{s}(0)\ |\ N(p)\!> & = &
  \bar{u}(p')
[f_V \gamma_{\lambda} + i \frac{f_M^s}{2m}
  \sigma_{\lambda\rho}q^\rho]\frac{1}{2} u(p).
\end{eqnarray}
As for the $q_{\mu}^2$ dependence of the form factors
we use the results of the latest analyses in
\cite{cvm,a_mass}:
\begin{eqnarray}
  f_V(q_{\mu}^2) &=& G_D(q^2_\mu)(1 + \mu_p \eta)(1 + \eta)^{-1},
  \label{eq:fv}\\
  f_M(q_{\mu}^2) &=& G_D(q^2_\mu)
   (\mu_p - \mu_n - 1 - \mu_n \eta)(1 + \eta)^{-1},\\ 
  f_A(q_{\mu}^2) &=& -g_A\ G_A(q^2_\mu)\label{eq_fa},\\
  f_P(q_{\mu}^2) &=& \frac{2m}{m^2_{\pi}-q_{\mu}^2}f_A(q_{\mu}^2),\\
  f_M^s(q_{\mu}^2) &=& G_D(q^2_\mu)
   (\mu_p + \mu_n - 1 + \mu_n \eta)(1 + \eta)^{-1},
   \label{eq:fsM}
\end{eqnarray}
with
\begin{eqnarray}
G_D(q^2_\mu) & = & \left(1 -  \frac{q^2_\mu}
{0.71\mbox{GeV}^2}\right)^{-2},\\
G_A(q^2_\mu) & = & \left(1 -  \frac{q^2_\mu}
{1.04\mbox{GeV}^2}\right)^{-2},
\label{eq:GA}
\end{eqnarray}
where $\mu_p=2.793$, $\mu_n = -1.913$, $\eta = -
\frac{q_\mu^2}{4m^2}$ and $m_{\pi}$ is the pion mass.
For $g_A$, we adopt the current standard value, 
$g_A$=1.267\cite{PDG}, instead of $g_A$=1.254 used in NSGK. 
In addition, as the axial-vector mass in Eq.(\ref{eq:GA}), 
we use the value which was obtained 
in the latest analysis\cite{a_mass} 
of (anti)neutrino scattering and 
charged-pion electroproduction.
The change in $G_A(q_{\mu}^2)$ is in fact
not consequential for \signud\ 
in the solar-$\nu$ energy region.
Regarding $f_P$, we assume PCAC and pion-pole dominance.
A contribution from this term is known to be proportional to the lepton
mass, which leads to very small contribution from the induced
pseudoscalar term in our case.
Although deviations from the naive pion-pole dominance 
of $f_P$ have been carefully studied\cite{fp}, 
we can safely neglect those deviations here.
For the IA current given above,
we carry out the non-relativistic reduction
in the same manner as in NSGK.

We now consider the exchange currents (EXC).
The axial-vector EXC, $A^{\mu}_{\rm EXC}$, 
consists of a pion-pole term and a non-pole term,
$\bar{A}^{\mu}_{\rm EXC}$. 
Using the PCAC hypothesis, however,
we can express $A^{\mu}_{\rm EXC}$ in terms of 
the non-pole contribution alone:
\begin{eqnarray}
  A^{\mu}_{\rm EXC} = \bar{A}^{\mu}_{\rm EXC} -
  \frac{q^{\mu}}{m^2_{\pi}-q_{\mu}^2} ( \bbox{q}\cdot
  \bbox{\bar{A}}_{\rm EXC} - \omega \bar{A}_{\rm EXC,0} ).
\label{eq_a_exc}
\end{eqnarray}
We therefore need only specify
a model for the non-pole terms;
the total contribution of $A^{\mu}_{\rm EXC}$
can be obtained with the use of Eq.(\ref{eq_a_exc}).
Regarding the space component of the axial-vector current, 
we employ \AEXC\ adjusted in such a manner
that the experimental value of \G3H\ be reproduced
(see the discussion in Introduction).
Following Schiavilla {\it et al.\ }\cite{model_I}, 
we consider the $\pi$-pair current 
(denoted by  $\pi S$), 
$\rho$-pair current ($\rho S$), 
$\pi$-exchange $\Delta$-excitation current 
($\Delta\pi$),
$\rho$-exchange $\Delta$-excitation current 
($\Delta\rho$) and 
$\pi\rho$-exchange current ($\pi\rho$).
The explicit expressions of these two-body currents
(acting on the $i$-th and $j$-th nucleons)
are as follows.
\footnotesize
\begin{eqnarray}
  \bar{\bbox{A}}^{\pm}_{ij}(\bbox{q}; \pi {\rm S}) &=& 
       - {f_A \over m} \, {f_{\pi N\!N}^2 \over m_\pi^2}
     \, { \bbox{\sigma}_j \cdot \bbox{k}_j 
     \over m_\pi^2 + \bbox{k}_j^2}
        f_\pi^2(\bbox{k}_j)
\left\{ (\bbox{\tau}_i\!\times\!\bbox{\tau}_j)^\pm
        \bbox{\sigma}_i\!\times\!\bbox{k}_j
        -\tau^\pm_j \left [ \bbox{q}+i
        \bbox{\sigma}_i\!\times\!( \bbox{p}_i + 
	\bbox{p}^\prime_i ) \right ]
        \right\}  + (i \rightleftharpoons j),\\
\nonumber\\
   \bar{\bbox{A}}^{\pm}_{ij}(\bbox{q}; \rho {\rm S}) &=&
      f_A {g_\rho^2 (1 + \kappa_\rho)^2 \over 4m^3}
       {f_\rho^2 (\bbox{k}_j) \over m_\rho^2 + \bbox{k}_j^2} 
       \left( \tau^\pm_j
       \left\{ 
       (\bbox{\sigma}_j\!\times\!\bbox{k}_j)\!\times \bbox{k}_j
    -  i \left[ \bbox{\sigma}_i\!\times\!
    (\bbox{\sigma}_j\!\times\!
       \bbox{k}_j ) \right]\!\times\!
       ( \bbox{p}_i + \bbox{p}^\prime_i) \right\}
                                               \right.  \nonumber\\
   &&  \left. + 
   (\bbox{\tau}_i\!\times\!\bbox{\tau}_j)^\pm \left\{ \bbox{q} 
       \bbox{\sigma}_i\!\cdot\! 
       ( \bbox{\sigma}_j\!\times\!\bbox{k}_j)
    +  i (\bbox{\sigma}_j\!\times\!\bbox{k}_j)\!\times\!
       (\bbox{p}_i\!+\!\bbox{p}^\prime_i) 
       - \left[ \bbox{\sigma}_i\!\times\!
       (\bbox{\sigma}_j\!\times\!\bbox{k}_j)
       \right]\!\times\!\bbox{k}_j \right\} \right) 
 + (i \rightleftharpoons j),\\
\nonumber\\
  \bar{\bbox{A}}^{\pm}_{ij} (\bbox{q};\Delta \pi)  &=& 
         {16 \over 25} \, f_A  \, {f_{\pi N\!N}^2 \over
           m_\pi^2 (m_\Delta - m)} \, { {\bbox \sigma}_j 
	   \cdot \bbox{k}_j
           \over m_\pi^2 + \bbox{k}_j^2} \, f_{\pi}^2 (\bbox{k}_j)
\left[ 4\, \tau^\pm_j \, \bbox{k}_j
     -({\bbox \tau}_i\!\times\!{\bbox \tau}_j)^\pm \,
     {\bbox \sigma}_i\!\times\!\bbox{k}_j  \right]
 + (i \rightleftharpoons j),
\label{eq:A1}\\
\nonumber\\
  \bar{\bbox{A}}^{\pm}_{ij} (\bbox{q};\Delta \rho)  &=&
        - {4 \over 25} \, f_A \, {g_\rho^2 (1 + \kappa_\rho)^2
         \over m^2 (m_\Delta - m)} \, {f_{\rho}^2(\bbox{k}_j) \over
          m_\rho^2 + \bbox{k}_j^2}
	\left\{ 4 \, \tau^\pm_j \,
         ( {\bbox \sigma}_j\!\times\!\bbox{k}_j) 
         \!\times\!\bbox{k}_j
         \right. 
     -( {\bbox \tau}_i \times {\bbox \tau}_j)^\pm \,
      {\bbox \sigma}_i\!\times\!
    \left [ ( {\bbox \sigma}_j\!\times\!\bbox{k}_j) 
      \!\times \bbox{k}_j \right ]\left. \right\}
\nonumber\\
&& + \,(i \rightleftharpoons j),
\label{eq:A2}\\
\nonumber\\
   \bar{\bbox{A}}^{\pm}_{ij}(\bbox{q}; \pi \rho) &=&
    2 f_A  {g_\rho^2 \over m} \, { \bbox{\sigma}_j
   \!\cdot\!\bbox{k}_j \over 
       (m_\rho^2 + \bbox{k}_i^2) (m_\pi^2 + \bbox{k}_j^2)}
       f_\rho (\bbox{k}_i) f_\pi (\bbox{k}_j) 
       (\bbox{\tau}_i\!\times\!\bbox{\tau}_j)^\pm
      \left[ (1\!+\!\kappa_\rho) 
      \bbox{\sigma}_i\!\times\!\bbox{k}_i  
      - i (\bbox{p}_i\!+\!\bbox{p}^\prime_i ) \right]
      + (i \rightleftharpoons j).
\end{eqnarray}
\normalsize
Here $m_\rho$ and $m_\Delta$ are the masses 
of the $\rho$-meson, and $\Delta$-particle, 
respectively;
$f_A$ is the axial form factor given in Eq.(\ref{eq_fa}).
The total three-momentum transfer is
$\bbox{q}\equiv\bbox{k}_i+\bbox{k}_j$,
with $\bbox{k}_{i(j)}$ being the momentum transferred
to the $i$-th ($j$-th) nucleon;
$\bbox{p}_i$ and $\bbox{p}^\prime_i$ are 
the initial and final momenta of the $i$-th nucleon.
The form factors, $f_\pi(\bbox{k})$ and 
$f_\rho(\bbox{k})$,
for the pion-nucleon and $\rho$-nucleon vertices
are parametrized as
\begin{eqnarray}
f_\pi(\bbox{k}) &=& 
{\Lambda_\pi^2 - m_\pi^2 \over 
\Lambda_\pi^2 + \bbox{k}^2}\ ,\quad\quad
f_\rho(\bbox{k}) =
{\Lambda_\rho^2 - m_\rho^2 \over 
\Lambda_\rho^2 + \bbox{k}^2}\ 
\label{eq:formfactor}
\end{eqnarray}
with $\Lambda_{\pi}=4.8$ fm$^{-1}$ 
and $\Lambda_{\rho}=6.8$ fm$^{-1}$.
The quark model has been used to relate
the coupling constants of the
$\pi N \Delta$, $\rho N \Delta$ 
and $A_\mu N\Delta$ vertices to the 
$\pi N\!N$, $\rho N\!N$, and $A_\mu NN$ vertices,
respectively.
Schiavilla {\it et al.\ }\cite{model_I} 
have pointed out
that the experimental value of \G3H\ can be 
reproduced if the strengths of 
$\bar{\bbox{A}}(\Delta\pi)$ in Eq.(\ref{eq:A1})
and 
$\bar{\bbox{A}}(\Delta\rho)$ in Eq.(\ref{eq:A2})
are reduced by a common factor of 0.8.
We employ here the same adjustment of 
$\bar{\bbox{A}}(\Delta\pi)$
and $\bar{\bbox{A}}(\Delta\rho)$.
For the third component
of the isovector current, 
we simply replace $\tau_i^{\pm}$ and 
$(\bbox{\tau}_i\times\bbox{\tau}_j)^{\pm}$ with 
$\tau_i^3/2$ and 
$(\bbox{\tau}_i \times\bbox{\tau}_j)^{3}/2$, respectively. 
(The same prescription will be applied 
to the other exchange currents as well.) 
For the time component we use the one-pion exchange
current (the so-called KDR current\cite{KDR}),
which gives the dominant EXC to $\bar{A}^{\pm}_{0\ ij}$.
The explicit form of the KDR current,
with a vertex form factor supplemented,\footnote{
For $A_0$ and the vector currents,
we use the same form factors as in NSGK.
They are parametrized as in Eq.(\ref{eq:formfactor}),
but the numerical values of 
$\Lambda_\pi$ and $\Lambda_\rho$ are:
$\Lambda_\pi=6.0\ {\rm fm}^{-1}$,
$\Lambda_\rho=7.3\ {\rm fm}^{-1}$} 
reads
\footnotesize
\begin{eqnarray}
\bar{A}^{\pm}_{0\ ij}(\bbox{q};KDR) &=& \frac{2}{if_A}
  \Biggl(\frac{f}{m_\pi}\Biggr)^2 
  f_{\pi}^2(\bbox{k}_j) 
\frac{\bbox{\sigma}_j \cdot \bbox{k}_j}
{m_\pi^2 + \bbox{k}_j^2}
(\bbox{\tau}_i \times \bbox{\tau}_j)^\pm 
+ \,(i \rightleftharpoons j)
\label{eq_KDR}.
\end{eqnarray}
\normalsize

Regarding the vector exchange currents,
we first note that the time component should be
negligibly small since its contribution 
vanishes in the static limit.
For the space component, $\bbox{V}$,
we take account of the pair, pionic, 
and isobar currents.
As in NSGK, we adopt the one-pion exchange model 
for the pair and pionic currents and 
the one-pion and one-$\rho$-meson exchange model 
for the isobar current. 
Their explicit expressions are:
\footnotesize
\begin{eqnarray}
  \bbox{V}^{\pm}_{ij}(\bbox{q};pair) &=& 
-2if_V \left( \frac{f}{m_ \pi} \right)^2
  f_{\pi}^2(\bbox{k}_j) 
\frac{\bbox{\sigma}_i (\bbox{\sigma}_j \cdot \bbox{k}_j)}
{m_\pi^2 + \bbox{k}_j^2}
(\bbox{\tau}_i \times \bbox{\tau}_j)^\pm
+ \,(i \rightleftharpoons j),\label{eq_pair}
\end{eqnarray}
\begin{eqnarray}
\bbox{V}^{\pm}_{ij}(\bbox{q};pionic) &=& 2i 
\left( \frac{f}{m_ \pi} \right)^2
  f_{\pi}(\bbox{k}_i) 
  f_{\pi}(\bbox{k}_j) 
\frac{ (\bbox{\sigma}_i \cdot 
\bbox{k}_i)(\bbox{\sigma}_j \cdot
\bbox{k}_j)(\bbox{k}_i-\bbox{k}_j )}
{(m_\pi^2 + \bbox{k}_i^2)(m_\pi^2 + \bbox{k}_j^2)}
(\bbox{\tau}_i \times \bbox{\tau}_j)^\pm,\\
\bbox{V}^{\pm}_{ij}(\bbox{q};\Delta) &=& 
- i 4 \pi \frac{f_V + f_M}{2 m}
\Biggl[  \frac{f_{\pi}^2(\bbox{k}_j)}
{m_\pi^2 + \bbox{k}_j^2} \bbox{q} \times\{ c_0 \bbox{k}_j
  \bbox{\tau}^\pm_j+d_1(\bbox{\sigma}_i
  \times \bbox{k}_j)(\bbox{\tau}_i
  \times \bbox{\tau}_j)^\pm \} 
(\bbox{\sigma}_j \cdot \bbox{k}_j)
  \nonumber \\
& & +\frac{f_{\rho}^2(\bbox{k}_j)}
{m_\rho^2 + \bbox{k}_j^2} \bbox{q} \times\left\{c_ \rho \bbox{k}_j \times
  (\bbox{\sigma}_j \times \bbox{k}_j)  
\bbox{\tau}^\pm_j+d_ \rho
  \bbox{\sigma}_i \times  (\bbox{k}_j \times 
  (\bbox{\sigma}_j \times \bbox{k}_j))(\bbox{\tau}_i
  \times \bbox{\tau}_j)^\pm\right\} \Biggr] 
\nonumber \\
& & + \,(i \rightleftharpoons j). \label{eq_delta}
\end{eqnarray}
\normalsize
The numerical values of the various parameters 
are
\footnotesize
\begin{eqnarray}
\frac{f^2}{4\pi}=0.08,\quad c_0m^3_\pi = 
0.188,\quad d_1m^3_\pi = -0.044,\quad
c_ \rho m^3_ \rho = 36.2,\quad  d_ \rho = 
- \textstyle{\frac{1}{4}}c_ \rho .
\end{eqnarray}
\normalsize
As discussed in NSGK, these values
lead to $np\rightarrow d\gamma$ cross sections
that agree with the experimental values.

Apart from the modifications explicitly mentioned above,
the theoretical framework of the present calculation
is the same as in NSGK \cite{NSGK} 
and, for further details of the formalism,
we refer to Sections II and III of NSGK

\vspace{4mm}
\noindent
{\bf 3.  Numerical results}

In reporting the numerical results of 
our calculation, we shall be primarily concerned
with the ``{\it standard case}",\footnote{
The {\it standard case} here should be distinguished
from the {\it standard run} in NSGK.}
which is characterized by the following features.
The calculational framework of 
the {\it standard case} is the same as in NSGK 
except for the specific points of improvements 
explained above.
Also, the numerical values of the input parameters
in the {\it standard case} are identical with those
used in the {\it standard run} in NSGK,
apart from the changes explicitly mentioned
in the preceding section.
As for the $NN$ interaction needed to generate
the initial and final nuclear wave functions,
the {\it standard case} employs 
the AV18 potential \cite{AV18}.\footnote{
The use of the AV18 potential here is 
consistent with the fact that, in \cite{pphep},
the strength of $\hat{d}^R$ was determined
using the 3-body wave functions obtained with 
the AV18 potential (and additional three-body forces). 
We also mention that the {\it standard run} in NSGK 
also uses the AV18 potential.}
In what follows, we largely concentrate on 
the {\it standard case} and discuss other cases
(to be specified as needed) only in the context
of assessing the model dependence.

For the {\it standard case}, we have calculated 
the total cross sections and differential cross sections
for the four reactions Eq.(\ref{eq_nu_CC})-(\ref{eq_bnu_NC}),
up to $E_\nu$ =170 MeV. 
In this communication, however, we concentrate 
on the quantities directly relevant 
to the SNO solar neutrino experiments
and limit ourselves to the neutrino reactions 
(both CC and NC)
for $E_\nu\le$ 20 MeV.\footnote{
A more extensive account of our calculation
will be published elsewhere.
The full presentation 
of the numerical results of the present work 
can be found at the web site:
$<$http://nuc003.psc.sc.edu/ 
$\tilde{ }\ $kubodera/NU-D-NSGK$>$.
\label{foot_web}}
The \signud\ corresponding 
to the {\it standard case}
is shown in Table \ref{tab_tot} 
as a function of $E_{\nu}$.\footnote{
The limited precision of our computer code
causes 0.1\% uncertainty in \signud\   
for $E_{\nu}\le 20$ MeV,
apart from uncertainties due to the model dependence 
to be discussed later in the text.}
The results given in Table \ref{tab_tot} 
should supersede the corresponding results in NSGK.
In the following we discuss comparison 
between the new and old estimates of 
\signud\ as well as
error estimates for the new calculation.

For clarity, when necessary, the total cross sections 
corresponding to the {\it standard case} 
of the present work are denoted by
\signudN, \signudCCN\  and \signudNCN;
those corresponding to
the {\it standard run} in NSGK are denoted by 
\signudNSGK, \signudCCNSGK\  and \signudNCNSGK.
The ratio of \signudCCN\  
to \signudCCNSGK\ is given 
for several representative values of $E_{\nu}$
in the first column of Table \ref{tab_tot2}.
Similar information for \signudNC\  is 
given in the second column.
As the table indicates,
\signudCCN\  is slightly larger than \signudCCNSGK;
the difference is $\sim 1.3\%$ for $E_{\nu}\sim 5$ MeV, 
$\sim 0.8\%$ for $E_{\nu}\sim 10$ MeV, 
and $\sim 0.4\%$ for $E_{\nu}\sim 20$ MeV.
A similar tendency is seen for \signudNC\  as well.
The origins of the difference between
\signudN\ and \signudNSGK\ will be analyzed below.

Changing the weak coupling constant
from $G_F$ to $G_F^{\,\,\prime}$
scales \signud\ by an overall factor of 
$(G_F^{\,\,\prime}/G_F)^2\sim 1.02$.
The effect of changing the value of $g_A$ 
can also be well simulated by an overall factor,
since the $\nu d$ reaction at low energies 
is dominated by the Gamow-Teller transition 
and hence \signud\  is essentially 
proportional to $g_A^2$. 
Thus the change of $g_A$ 
from $g_A$ = 1.254 to $g_A$ = 1.267
enhances \signud\  in the low-energy region
by another factor of 
$(1.267/1.254)^2\sim 1.02$.

\begin{figure}[h]
\caption{Contributions of \AEXC\ to \signudCC;
$\xi$ defined in the text is plotted
for Model I (solid line) and Model II (dashed line). 
The dash-dotted line represents
the ``normalized" version of Model II
described in the text.}
\begin{center}
 \epsfig{file=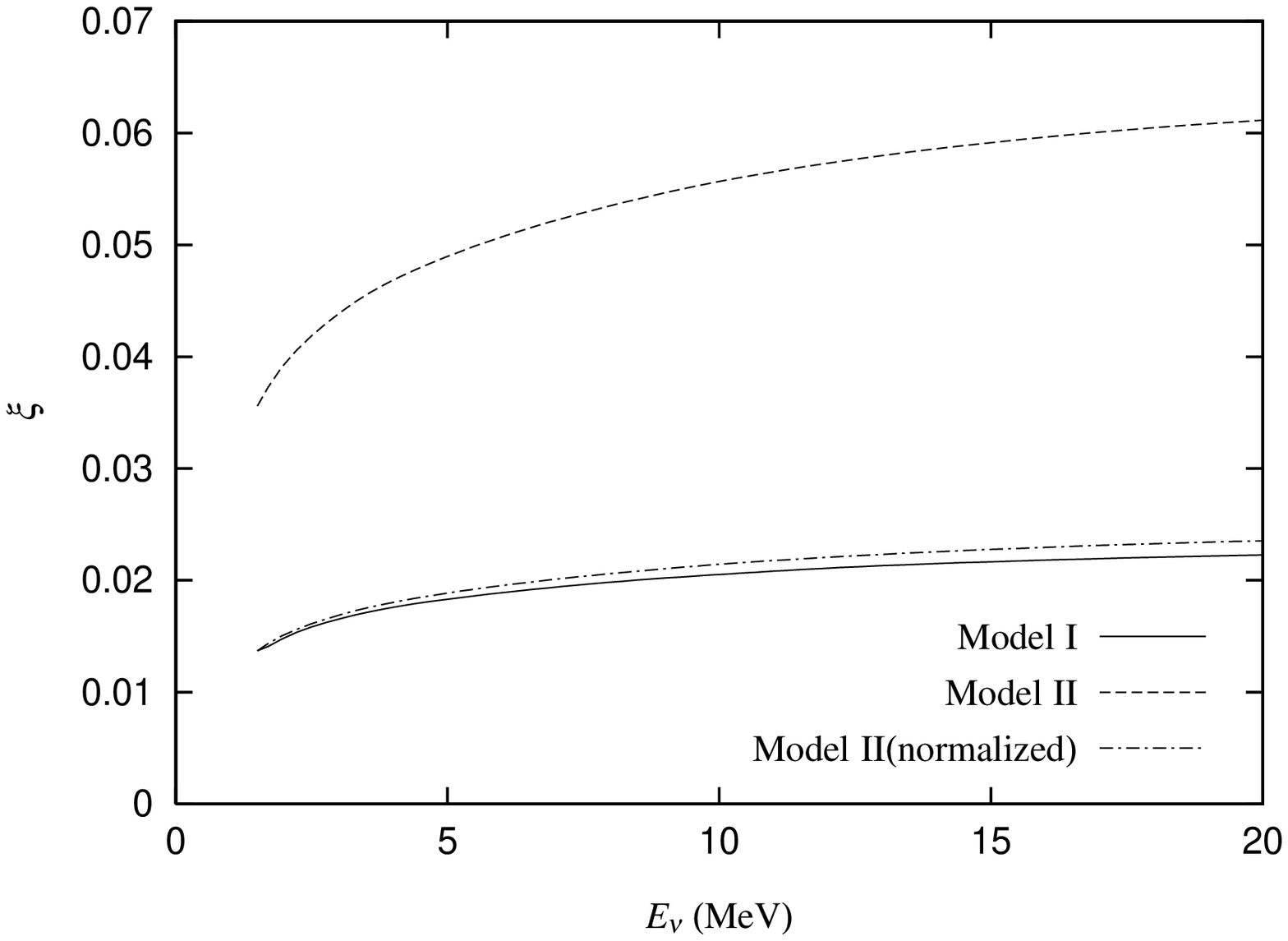,scale=0.6} 
\end{center}
\label{fig_exc}
\end{figure}

In discussing the consequences
of the change in \AEXC, 
it is convenient to introduce the terms,
Models I and II. 
As described earlier,
the {\it standard case} in the present calculation
uses \AEXC\  given in \cite{model_I}
and recapitulated in the preceding section.
We refer to this choice of \AEXC\  as Model I.
Meanwhile, \AEXC\  used in the 
{\it standard run} in NSGK consists of 
$\bbox{A}(\Delta\pi)$ and 
$\bbox{A}(\Delta\rho)$ alone, 
and its strength is adjusted to reproduce 
the $np\rightarrow \gamma d$ rate. 
We refer to this choice of \AEXC\ as Model II.
For each of Models I and II, 
Table \ref{tab_add_exc} gives 
the contributions from the individual terms
in $A_\mu$ as well as that from $V_\mu$, 
the vector current.
The table indicates
that in either case the corrections to the IA values
are dominated by the contributions
from $\bbox{A}(\Delta\pi)$ 
and 
$\bbox{A}(\Delta\rho)$.
To facilitate further comparison 
between Models I and II,
we consider the ratio, $\xi$, defined by 
$\xi \equiv
\ [\sigma_{\nu d}({\rm IA}\!+\!\bbox{A}_{\rm EXC})
-\sigma_{\nu d}({\rm IA})]
/\sigma_{\nu d}({\rm IA})$.
Here, $\sigma_{\nu d}({\rm IA})$ is 
the result obtained with the IA current alone, while 
$\sigma_{\nu d}({\rm IA}\!+\!\bbox{A}_{\rm EXC})$
represents the result obtained 
with the IA current plus \AEXC.
Fig.\ \ref{fig_exc} gives $\xi$ for the CC reaction
as a function of $E_{\nu}$.
The solid line shows $\xi$ for Model I, while
the dashed line gives $\xi$ for Model II.
It is seen that the contribution 
of \AEXC\  in Model I is smaller
than that in Model II by $2\sim 4\%$.
This difference is mainly due to 
the reduced strength of the $\Delta$-excitation
currents in Model I.
For further discussion, we normalize 
$\xi${\scriptsize{(Model II)}}
for the CC reaction
by an overall multiplicative factor
chosen in such a manner that
the normalized 
$\xi${\scriptsize{(Model II)}}
reproduces 
$\xi${\scriptsize{(Model I)}}
at the reaction threshold.
This normalized result is given by   
the dash-dotted line in Fig.\ref{fig_exc}.
We observe that 
the dash-dotted line exhibits a slight deviation 
from the solid line (Model I).
This deviation reflects a slight difference 
in the $E_\nu$-dependences of the \AEXC\ contribution
for Models I and II.
The main cause of this difference can be traced
as follows. 
From Table \ref{tab_add_exc} 
we can deduce that the dominant contributions 
coming from the $\Delta$-excitation currents 
have almost the same $E_\nu$-dependence
for Models I and II,
although their absolute values 
differ for the two models. 
On the other hand, the contributions of
$\bbox{A}(\pi S)$, $\bbox{A}(\rho S)$ 
and $\bbox{A}(\pi\rho)$,
which are included only in Model I,
are virtually $E_\nu$-independent,
and their magnitudes are small. 
These features lead to the 
slighly weaker $E_\nu$-dependence in \AEXC(Model I)
than in \AEXC(Model II).
The behavior of $\xi$ for the NC reaction (not shown)
is similar to $\xi$ for the CC reaction.

The error estimate adopted in NSGK
essentially consists in taking 
the difference between Models I and II 
as a typical measure of model dependence.
As mentioned in the introduction, however,
Model II, which fails to explain \G3H,
should not be given the same status as Model I.
To attain a more reasonable estimate of
the theoretical uncertainty, 
we propose the following interpretation 
of the feature seen in Fig.\ \ref{fig_exc}.
The fact that Model I has been adjusted 
to reproduce \G3H 
means that it can yield model-independent results
at a specific kinematics 
but that, without additional experimental information,
the $E_\nu$-dependence of \signud\ 
cannot be fully controlled. 
This uncertainty may be assessed 
from the difference between the solid
and dash-dotted lines in Fig.\ref{fig_exc}. From this argument we assign
0.2\% uncertainty  
to the contribution of \AEXC\ 
to \signud\  in the solar neutrino energy range, 
$E_\nu<$ 20 MeV.

We recapitulate our discussion regarding
the change from \signudNSGK\ to \signudN:
a $\sim$4\% enhancement of \signud\  due
to the changes in  the Fermi constant and $g_A$
and a $\sim$3\% reduction due to the use 
of the new \AEXC\ (Model I) that reproduces \G3H. 
These two changes partially cancelling each other,
the net result is 
the enhancement of \signud\ by $\sim$1\%,
and this is what is seen in Table \ref{tab_tot2}.

As mentioned earlier,
an additional important measure of reliability 
of our SNPA calculation is obtained by
comparing it with the results of
an EFT* calculation by Ando {\it et al.}\ \cite{ando}. 
By using the value of the low-energy constant, 
$\hat{d}^R$, fixed to reproduce the experimental value 
of \G3H \cite{pphep},
Ando {\it et al.}\ \cite{ando} have carried out 
a parameter-free EFT-motivated calculation of \signud. 
Although the cut-off regularization method 
used in \cite{ando} can introduce
the cut-off dependence into the formalism,
it has been checked \cite{ando}
that this dependence is negligibly small
for a physically reasonable range
of the cut-off parameter;
the relative variation in \signud\   
only amounts to 0.02\%,
which is much smaller than
the above-mentioned 0.2\% uncertainty
inherent in our SNPA calculation.
In fact, the uncertainty in \signud\ 
obtained by Ando {\it et al.}\ 
is dominated by the 0.5\% error
resulting from the uncertainty in
the experimental value of \G3H.
We now compare \signudN\  
with $\sigma_{\nu d}${\scriptsize{(EFT*)}}
obtained in the EFT* calculation of 
Ando {\it et al.} \cite{ando}.
Since Ref.\cite{ando} only includes
the s-wave of the final $NN$ state,
we compare $\sigma_{\nu d}${\scriptsize{(EFT*)}} 
with $\sigma_{\nu d}$($s$-wave),
which represents the $s$-wave contribution
to \signud\ calculated for the
{\it standard case}.
The ratio, 
$\eta\equiv \sigma_{\nu d}
{\mbox {\scriptsize{(EFT*)}}}
/\sigma_{\nu d}$($s$-wave),
is shown in Table \ref{tab_rat_eft}, 
from which we can conclude that SNPA and EFT
give identical results at the 1\% level.

We proceed to consider the $NN$ potential dependence. 
As mentioned, the CD-Bonn potential
is somewhat distinct from the potentials
considered in NSGK,
in that it has a significantly smaller
$D$-state probability;
$P_D${\scriptsize{(CD-Bonn)}} = 4.2\%
as compared with $P_D${\scriptsize{(AV18)}} = 5.8\%.
We compare in Table \ref{tab_NN}
the \signudNC\ obtained 
with the AV18 and CD-Bonn potentials. 
The difference between the two cases 
is found to be practically negligible.
With the CD-Bonn potential, 
because of its larger $S$-state probability,
the contribution from the IA current 
becomes larger,
whereas the contribution from \AEXC\  
is smaller due to its reduced $D$-state probability. 
Our explicit calculation
demonstrates that the cancellation
between these two opposing tendencies
is almost perfect, providing a yet another
manifestation of the robustness
of the calculated \signud.
A similar stabilizing mechanism was
noticed by Schiavilla {\it et al.} 
in their study of the $pp$-fusion 
cross section \cite{model_I}.

Apart from the absolute values of 
\signudCC and \signudNC, the ratio, 
$R \equiv \sigma_{\nu d}^{NC}
/\sigma_{\nu d}^{CC}$, 
is also an important quantity for SNO experiments.
As emphasized by Bahcall and Lisi \cite{BL96},
a measurement at SNO of the ratio 
of the number of the NC to CC events
(the NC/CC ratio) would place stringent
constraints on various neutrino oscillation scenarios.
Since the precision of the predicted value 
of the NC/CC ratio is affected by the uncertainty
in $R$ (see Fig.7 in \cite{BL96}),
we discuss the model dependence of $R$.
Table \ref{tab_rat} gives the values of $R$ 
calculated for the various cases discussed above.
The table indicates that the model dependence
of $R$ is smaller than 
that of \signudCC\  and \signudNC\  themselves.
The simple IA calculation gives 
a value of $R$ that agrees with $R_{stnd}$ 
within 0.2\%. 
The variance between the {\it standard case}
and Model II is less than 0.5\% --
it is to be recalled that Model II is 
a rather extreme case.
Furthermore, the difference between 
the {\it standard case}
and the EFT* calculation does not exceed 0.4\%.
We therefore consider it reasonable to assign
0.5\% accuracy to $R$.
This is an improvement by a factor of $\sim$2
over the precision reported in \cite{KN,NSGK}.
 
The stability of $R$ 
can be understood as follows.
We first note the following two features.
(1) The contribution of the isoscalar current,
which only participates in the NC reaction,
is negligibly small in our case;
(2) Although the iso-vector 
vector and axial-vector currents enter
in different ways into the nuclear currents 
responsible for the CC or NC reactions,
the contribution of the vector current 
is much smaller than that
of the axial-vector current
in the solar neutrino energy regime.
As a consequence of these two facts,
the transition operators
for the NC and CC reactions in the present case
are, to good accuracy, related 
by a rotation in isospin space.
So, if there were no isospin breaking effects
in the nuclear wave functions,
the CC and NC transition amplitudes 
would be simply related by the
Wigner-Eckart theorem in isospin space,
leading to the complete model independence of $R$.
In reality, there are isospin-breaking effects
in the two-nucleon wave functions, 
but these ``{\it external effects}"
are expected to be under good control 
so long as one uses high-quality $NN$
potentials that reproduce the $NN$ data accurately.

\vspace{5mm}
\noindent
{\bf 4. Discussion and Summary}

\vspace{3mm}

Although we do not directly address
the issue of radiative corrections (RC) here,
we make a few remarks on it.
RC can affect \signud\  at the level of a few percents. 
According to Kurylov {\it et al.}\ \cite{RC_KRV},
RC increases \signudCC\ by $4\%$ at low $E_\nu$
and by $3\%$ at the higher end 
of the solar neutrino energy,
while RC leads to an $E_\nu$-independent
increase of \signudNC\  by $\sim 1.5\%$. 
The RC for \signudCC\  consists of 
the ``inner'' and ``outer'' corrections. 
The former is sensitive to hadronic dynamics
but energy-independent, 
while the latter is largely independent of
hadronic dynamics but has energy-dependence.
The use of experimental value of $G_F^{\,\,\prime}$
\cite{TH} obtained from $0^+ \rightarrow 0^+$
nuclear $\beta$-decays allows one 
to take account of the bulk of the ``inner'' corrections. 
To obtain reasonable up-to-date estimates
of the remaining ``outer" corrections,
we may proceed as follows.
For \signudCC,
we may adopt as the ``outer" correction
the  difference between the result 
of Kurylov {\it et al.}\ ($4\%$ - $3\%$) 
and the estimated ``inner'' corrections ($2.4\%$).
For \signudNC, there is no ``outer" corrections
at the level of precision of this article.
In adopting this prescription, 
we are leaving unaddressed a delicate issue of 
the possible difference between RC for the single nucleon
and RC for multi-nucleon systems,
but this seems to be the best 
we can do at present.

To summarize, 
we have improved NSGK's calculation \cite{NSGK}
for the $\nu d$ reactions
by updating some of its inputs
and with the use of 
the axial-vector exchange current
the strength of which is controlled by \G3H. 
We have also taken into account 
the results of a recent parameter-free
EFT* calculation \cite{ando}.
The new value of \signud, denoted by \signudN,
is slightly larger than 
$\sigma_{\nu d}${\scriptsize{(NSGK)}}
reported in \cite{NSGK};  
$\sigma_{\nu d}{\mbox{\scriptsize{(Netal)}}}
/\sigma_{\nu d}{\mbox{\scriptsize{(NSGK)}}}
\sim 1.01$.
Based on the arguments presented above,
we consider it reasonable to assign
1\% uncertainty to \signudN\
given in Table \ref{tab_tot},
and $\sim$0.5\% uncertainty to $R$. 
The results in Table \ref{tab_tot}, however,
do not include radiative corrections
except for those already
incorporated into the empirical value of 
$G_F^{\,\,\prime}$,
which subsumes the bulk of the inner radiative 
corrections for nuclear $\beta$-decay.
With the inclusion of 
the remaining radiative corrections, 
\signudCC\ is likely to become larger
than \signudCCN\  by up to $\sim$2 \%,  
while \signudNC\  is expected to
lie within the quoted $\sim$1 \% error
of \signudNCN.
 
\begin{center}
{\footnotesize\bf{ACKNOWLEDGMENTS}}\\
\end{center}
We are grateful to J. Beacom for his useful criticism
on Ref.\cite{NSGK} and for calling
our attention to the importance of the radiative
corrections. 
Thanks are also due to John Bahcall 
for his interest in and useful comments on this work.
We are deeply indebted to R. Machleidt
for his generosity in allowing us to use 
his CD-Bonn potential code.
This work is supported in part by the US National
Science Foundation, Grant No. PHY-9900756 and
No. INT-9730847, and also by the Japan Society for 
the Promotion of Science, Grant No. (c) 12640273.

\begin{table}[h]
\caption{Calculated values of \signudCC \ and 
\signudNC \ in units of cm$^2$. 
The ``-$x$'' in the parentheses 
means 10$^{-x}$; 
thus an entry like 4.579(-48) stands for 
$4.579\times 10^{-48}\ {\rm cm}^2$.}
\footnotesize
\begin{center}
 \begin{tabular}[t]{rccrcc}\hline
 $E_{\nu}\ \ $&&&$E_{\nu}\ \ $&&\\
 (MeV) & $\nu_e d\rightarrow e^-pp$& $\nu d\rightarrow \nu pn$&
 (MeV) & $\nu_e d\rightarrow e^-pp$& $\nu d\rightarrow \nu pn$\\ \hline
 1.5&4.680 (-48)&0.000 (  0)& 8.8&1.911 (-42)&7.530 (-43)\\
 1.6&1.147 (-46)&0.000 (  0)& 9.0&2.034 (-42)&8.070 (-43)\\
 1.8&1.147 (-45)&0.000 (  0)& 9.2&2.160 (-42)&8.629 (-43)\\
 2.0&3.670 (-45)&0.000 (  0)& 9.4&2.291 (-42)&9.209 (-43)\\
 2.2&7.973 (-45)&0.000 (  0)& 9.6&2.425 (-42)&9.809 (-43)\\
 2.4&1.428 (-44)&4.346 (-47)& 9.8&2.565 (-42)&1.043 (-42)\\
 2.6&2.279 (-44)&4.322 (-46)&10.0&2.708 (-42)&1.107 (-42)\\
 2.8&3.369 (-44)&1.478 (-45)&10.2&2.856 (-42)&1.173 (-42)\\
 3.0&4.712 (-44)&3.402 (-45)&10.4&3.007 (-42)&1.241 (-42)\\
 3.2&6.324 (-44)&6.372 (-45)&10.6&3.164 (-42)&1.311 (-42)\\
 3.4&8.216 (-44)&1.052 (-44)&10.8&3.324 (-42)&1.383 (-42)\\
 3.6&1.040 (-43)&1.594 (-44)&11.0&3.489 (-42)&1.458 (-42)\\
 3.8&1.289 (-43)&2.274 (-44)&11.2&3.658 (-42)&1.534 (-42)\\
 4.0&1.569 (-43)&3.098 (-44)&11.4&3.832 (-42)&1.612 (-42)\\
 4.2&1.881 (-43)&4.072 (-44)&11.6&4.010 (-42)&1.693 (-42)\\
 4.4&2.225 (-43)&5.202 (-44)&11.8&4.192 (-42)&1.775 (-42)\\
 4.6&2.604 (-43)&6.492 (-44)&12.0&4.379 (-42)&1.860 (-42)\\
 4.8&3.016 (-43)&7.947 (-44)&12.2&4.570 (-42)&1.947 (-42)\\
 5.0&3.463 (-43)&9.570 (-44)&12.4&4.766 (-42)&2.035 (-42)\\
 5.2&3.945 (-43)&1.136 (-43)&12.6&4.966 (-42)&2.126 (-42)\\
 5.4&4.463 (-43)&1.333 (-43)&12.8&5.171 (-42)&2.219 (-42)\\
 5.6&5.017 (-43)&1.548 (-43)&13.0&5.380 (-42)&2.314 (-42)\\
 5.8&5.608 (-43)&1.780 (-43)&13.5&5.923 (-42)&2.561 (-42)\\
 6.0&6.236 (-43)&2.031 (-43)&14.0&6.495 (-42)&2.822 (-42)\\
 6.2&6.902 (-43)&2.300 (-43)&14.5&7.095 (-42)&3.095 (-42)\\
 6.4&7.605 (-43)&2.587 (-43)&15.0&7.724 (-42)&3.382 (-42)\\
 6.6&8.347 (-43)&2.894 (-43)&15.5&8.383 (-42)&3.682 (-42)\\
 6.8&9.127 (-43)&3.219 (-43)&16.0&9.071 (-42)&3.995 (-42)\\
 7.0&9.946 (-43)&3.562 (-43)&16.5&9.789 (-42)&4.323 (-42)\\
 7.2&1.080 (-42)&3.925 (-43)&17.0&1.054 (-41)&4.663 (-42)\\
 7.4&1.170 (-42)&4.308 (-43)&17.5&1.131 (-41)&5.017 (-42)\\
 7.6&1.264 (-42)&4.709 (-43)&18.0&1.212 (-41)&5.385 (-42)\\
 7.8&1.362 (-42)&5.130 (-43)&18.5&1.296 (-41)&5.767 (-42)\\
 8.0&1.464 (-42)&5.571 (-43)&19.0&1.383 (-41)&6.162 (-42)\\
 8.2&1.569 (-42)&6.031 (-43)&19.5&1.474 (-41)&6.571 (-42)\\
 8.4&1.679 (-42)&6.511 (-43)&20.0&1.567 (-41)&6.994 (-42)\\
 8.6&1.793 (-42)&7.010 (-43)&&&\\
 \\\hline
 \end{tabular}
\end{center}
\normalsize
\label{tab_tot}
\end{table}
\begin{table}[h]
\caption{Comparison of the present results
with those of NSGK \cite{NSGK}.
The ratio, 
$\sigma_{\nu d}{\mbox{\scriptsize{(Netal)}}}/
 \sigma_{\nu d}{\mbox{\scriptsize{(NSGK)}}}$,
is given for representative values 
of $E_\nu$. }
\footnotesize
\begin{center}
  \begin{tabular}[t]{ccc}\hline
 $E_{\nu}$ (MeV)
 & $\nu_e d\rightarrow e^-pp$
 & $\nu d\rightarrow \nu pn$\\ \hline
  5& 1.013& 1.011\\
 10& 1.008& 1.006\\
 15& 1.006& 1.003\\
 20& 1.004& 1.001\\\hline
 \end{tabular}
\end{center}
\normalsize
\label{tab_tot2}
\end{table}
\begin{table}[h]
\caption{For Models I and II are shown
the cumulative contributions to \signudCC\
from the various components in the current.
The row labeled ``IA" gives \signudCC\  obtained
with the IA currents in $A_\mu$ and $V_\mu$, 
and the next row labeled ``+$\bbox{V}_{\rm EXC}$" 
gives \signudCC\ 
that includes the contributions of  
the IA currents and $\bbox{V}_{\rm EXC}$,
the exchange current in $\bbox{V}$.
Similarly, an entry in the $n$-th row
(counting from the row labeled ``IA'')
includes the coherent contributions
of all the currents listed in the first $n$ rows.
The numbers in the last row are obtained with
the full currents.
The parenthesized number in the $n$-th row
gives the ratio, 
$\sigma_{\nu d}^{CC}\!$($n$-th row)/
$\sigma_{\nu d}^{CC}\!$(($n$-1)-th row),
which represents a factor 
by which \signudCC\ changes
when the new term is added.}
\vspace{5mm}
\footnotesize
\begin{center}
 \begin{tabular}[t]{lllll}
 \multicolumn{5}{c}
 {\signudCC\,\, 
 ($\times 10^{-42}{\rm cm}^2$)}\\ \hline
 \multicolumn{5}{c}{Model I}\\ \hline
 $E_{\nu}$ &5 MeV&10 MeV&15 MeV&20 MeV \\ \hline
 \ \ IA            & 0.3397 (  -  )&  2.646 (  -  )&  7.526 (  -  )&  15.23 (  -  )\\
 +$\bbox{V}_{\rm EXC}$ & 0.3401 (1.001)&  2.654 (1.003)&  7.560 (1.005)&  15.33 (1.006)\\
 +$\pi\Delta$      & 0.3474 (1.022)&  2.719 (1.025)&  7.758 (1.026)&  15.74 (1.027)\\
 +$\rho\Delta$     & 0.3448 (0.992)&  2.695 (0.991)&  7.687 (0.991)&  15.59 (0.991)\\
 +$\pi S$          & 0.3456 (1.002)&  2.702 (1.003)&  7.707 (1.003)&  15.63 (1.003)\\
 +$\rho S$         & 0.3447 (0.997)&  2.694 (0.997)&  7.682 (0.997)&  15.58 (0.997)\\
 +$\pi-\rho$       & 0.3463 (1.005)&  2.708 (1.005)&  7.724 (1.005)&  15.67 (1.006)\\
 +$A^0_{\rm EXC}$  & 0.3463 (1.000)&  2.708 (1.000)&  7.724 (1.000)&  15.67 (1.000)\\\hline
 \end{tabular}

 \vspace{5mm}
 \begin{tabular}[t]{lllll}\hline
 \multicolumn{5}{c}{Model II}\\ \hline
 $E_{\nu}$ &5 MeV&10 MeV&15 MeV&20 MeV \\ \hline
 \ \ IA            & 0.3397 (  -  )&  2.646 (  -  )&  7.526 (  -  )&  15.23 (  -  )\\
 +$\bbox{V}_{\rm EXC}$ & 0.3401 (1.001)&  2.654 (1.003)&  7.560 (1.005)&  15.33 (1.006)\\
 +$\pi\Delta$      & 0.3612 (1.062)&  2.841 (1.071)&  8.128 (1.075)&  16.52 (1.078)\\
 +$\rho\Delta$     & 0.3567 (0.988)&  2.801 (0.986)&  8.007 (0.985)&  16.26 (0.985)\\
 +$A^0_{\rm EXC}$  & 0.3567 (1.000)&  2.801 (1.000)&  8.008 (1.000)&  16.26 (1.000)\\\hline
 \end{tabular}\\
\end{center}
\normalsize
\label{tab_add_exc}
\end{table}
\begin{table}[h]
\caption{Comparison of SNPA and EFT calculations.
The ratio, $\eta\equiv \sigma_{\nu d}
{\mbox {\scriptsize{(EFT*)}}}
/\sigma_{\nu d}(s{\rm -wave})$,
is given for representative values of $E_\nu$.}
\footnotesize
\begin{center}
  \begin{tabular}[t]{ccc}\hline
 $E_{\nu}$ (MeV) 
 & $\nu_e d\rightarrow e^-pp$
 & $\nu d\rightarrow \nu pn$\\ \hline
 5& 1.003& 1.004\\
 10& 1.001& 1.003\\
 15& 0.999& 1.002\\
 20& 0.998& 1.001\\\hline
 \end{tabular}
\end{center}
\normalsize
\label{tab_rat_eft}
\end{table}

\begin{table}
\caption{Dependence of \signudNC\ on $NN$ potentials. 
`Bonn' and `AV18' represent 
the results obtained with the CD-Bonn and
the AV18 potentials, respectively.}
\footnotesize
\begin{center}
  \begin{tabular}[t]{crrrrrrrr}
 \multicolumn{9}{c}{\signudNC\ \,\,($\times 10^{-42}
 {\rm cm}^2$)}\\ \hline
 $E_{\nu}$ &\multicolumn{2}{c}{5 MeV}&\multicolumn{2}{c}{10
  MeV}&\multicolumn{2}{c}{15 MeV}&\multicolumn{2}{c}{20 MeV}\\
 &Bonn&AV18&Bonn&AV18&Bonn&AV18&Bonn&AV18\\ \hline
 IA         & 0.09459& 0.09390&  1.091&  1.083&  3.327&  3.300&  6.871&  6.814\\
 IA+EXC     & 0.09557& 0.09570&  1.104&  1.107&  3.373&  3.382&  6.973&  6.994\\
 \hline
 \end{tabular}
\end{center}
\normalsize
\label{tab_NN}
\end{table}

\begin{table}
\caption{The ratio,
$R\equiv \sigma(\nu d \rightarrow \nu np)
/\sigma(\nu_e d\rightarrow e^-pp)$,
calculated for representative values of $E_\nu$.
The second column gives $R_{stnd}$ corresponding 
to the {\it standard case}. 
The third, fourth and fifth columns give 
$R_{\rm IA}/R_{stnd}$,
$R_{\rm [Model\,\, II]}/R_{stnd}$
and 
$R_{\rm [EFT*]}/R_{stnd}$, respectively.
Here $R_{\rm IA}$ corresponds to $R$
obtained with the IA current alone,
while $R_{\rm [Model\,\, II]}$ and $R_{\rm [EFT*]}$
corresponds to Model II and EFT*, respectively.}
\footnotesize
\begin{center}
  \begin{tabular}[t]{ccccc}\hline
 $E_{\nu}$ (MeV) 
 &$R_{stnd}$&  IA  & Model II& EFT*\\ \hline
   5&   0.2764& 1.000& 1.004& 1.001\\
  10&   0.4087& 1.001& 1.004& 1.002\\
  15&   0.4378& 1.002& 1.004& 1.004\\
  20&   0.4464& 1.002& 1.005& 1.004\\
 \hline
 \end{tabular}
\end{center}
\normalsize
\label{tab_rat}
\end{table}

\end{document}